\documentstyle[12pt,epsfig]{article}
\newcommand{\be}{\begin{equation}}
\newcommand{\ee}{\end{equation}}
\newcommand{\ba}{\begin{eqnarray}}
\newcommand{\ea}{\end{eqnarray}}
\catcode`\^^I=13\def^^I{\ \ \ \ \ \ \ \ }
\begin{document}
\begin{flushright}
Revised on \today
\end{flushright}
\vspace{.5in}
\begin{center}
{\Huge The Relaxation Method for Solving
the Schr\"{o}dinger Equation in Configuration Space
with the Coulomb and Linear Potentials}\\
\vspace{.25in}
{\Large Alfred Tang, Daniel R. Shillinglaw and George Nill}\\
\vspace{.15in}
{\em Physics Department, University of Wisconsin - Milwaukee,}\\
{\em P. O. Box 413, Milwaukee, WI 53201.}\\
Emails: atang@uwm.edu, drshilli@uwm.edu, geonill@uwm.edu
\vspace{5mm} \noindent

\noindent
{\bf Abstract}
\end{center}
\noindent
The non-relativistic
Schr\"{o}dinger equation with the linear and Coulomb potentials is solved 
numerically in configuration space using the relaxation method.  The numerical
method presented in this paper is a plain explicit Schr\"{o}dinger solver
which is conceptually simple and is suitable for advanced undergraduate 
research.

\newpage

\section{Introduction}
This work came out of an undergraduate research project at the University 
of Wisconsin-Milwaukee.  The analytical solution of the hydrogen atom is a
typical topic in advanced undergraduate or beginning graduate quantum 
mechanics which utilizes the series expansion technique.  The analytic solution
does not only illustrate the mathematical approach of solving differential
equations, it also provides a benchmark for testing numerical methods.  
Initially the numerical solution of the hydrogen atom was intended to offer 
the undergraduate students in our department (the second and third authors of 
this paper) an opportunity to learn the elements of scientific computation and 
basic research strategies.  The students began by learning the theory of 
partial differential equations, linux programming in $C$, and basic numerical 
methods.  In order to give the students a taste of original theoretical 
research, they were asked to check the eigenvalues generated by a Nystrom 
momentum space code from a new paper with the configuration space code 
which they helped to develop. 
This paper documents the $r$-space code for the numerical solution of the 
non-relativistic Schr\"{o}dinger equation (NRSE) with the Coulomb and linear 
potentials.  It is hoped that this work is helpful to those students and 
teachers who want to integrate numerical methods with a traditional quantum 
mechanics curriculum.

\section{Schr\"{o}dinger Equation in configuration Space}
The basis of the wavefunction of the Sch\"{o}dinger equation in configuration
space is taken to be
\be
\phi_{lm}({\mathbf r})={R_{l}(r)\over r}\,Y_{lm}(\Omega)
\ee
The radial part of the equation for a hydrogen atom is~\cite{greiner}
\be
{d^{2}R_{l}\over dr^{2}}+\left[{2\mu\over\hbar}\left(E+{e^{2}\over r}\right)
-{l(l+1)\over r^{2}}\right]R_{l}=0.
\label{se}
\ee
A computer cannot integrate $r$ from zero to infinity.  Therefore
we must map $[0,\infty)\to [0,1]$ by
\be
r={x\over 1-x}
\ee
It implies that 
\be
{dx\over dr}=(1-x)^{2}.
\ee
From now on, a new symbol for the radial wavefunction will be used, {\em i.e.} 
$y\equiv R_{l}$.  To transform the
Schr\"{o}dinger equation to the new space, we first transform the second
derivative,
\ba
{d^{2}y\over dr^{2}}&=&{dx\over dr}{d\over dx}\left({dx\over dr}
{dy\over dx}\right)\nonumber\\
&=&(1-x)^{2}{d\over dx}\left((1-x)^{2}{dy\over dx}\right)\nonumber\\
&=&(1-x)^{2}\left[2(1-x)\,{dy\over dx}+(1-x)^{2}{d^{2}y\over dx^{2}}\right]
\label{2der}
\ea
By substituting Eq.~[\ref{2der}] into Eq.~[\ref{se}] and letting
$\hbar\to 1$ and
\be
e^{2}={1\over 137},\qquad \mu=0.5107208\times 10^{6}\,eV,
\ee
in natural units, the Schr\"{o}dinger equation is transformed as
\be
{d^{2}y\over dx^{2}}+{2\over 1-x}\,{dy\over dx}+{1\over (1-x)^{4}}
\left[2\mu\left(E+{1-x\over x}\,e^{2}\right)-\left({1-x\over x}\right)^{2}
l(l+1)\right]y=0.
\label{de_h2}
\ee
With the benefit of hindsight, we know that the non-zero portion of the 
solution of Eq.~[\ref{de_h2}]
will be tightly bound to a narrow margin near $x\sim 0$.  In order to improve
the numerical accuracy and the quality of the plots, we like to focus on 
the non-zero portion of the solutions.  A new configuration variable $z$ and
a redefinition of $x$
\be
z\equiv {r\over a_{0}},\qquad z={x\over 1-x},
\label{redef1}
\ee
where
\be
a_{0}\equiv {\hbar^{2}\over\mu e^{2}}=2.6831879\times 10^{-4}\,eV^{-1},
\label{redef2}
\ee
are used to zoom into the small $x$ region.  Eq.~[\ref{de_h2}] is modified as
\be
{d^{2}y\over dx^{2}}+{2\over 1-x}\,{dy\over dx}+{1\over (1-x)^{4}}
\left[2\mu\,a_{0}^{2}\left(E+{1-x\over x}\,{e^{2}\over a_{0}}\right)-
\left({1-x\over x}\right)^{2}l(l+1)\right]y=0.
\label{de_a0}
\ee

\section{Relaxation Method}
\label{der}
The shooting method typically shoots from one boundary point to another
using Runga-Kutta integration.  In the case of the
Schr\"{o}dinger equation, the boundary points at $x=0$ and $x=1$ are both
singular.  A one-point shoot will not converge when the code marches toward
a singularity.  A two-point shoot marches from both boundary points 
at $x=0$ and $x=1$ to match a third boundary point somwhere in between.
Even if it works, two-point shoot requires too much {\it a priori} knowledge 
of the wavefunction.  Furthermore any Runga-Kutta based alogrithms, such as
the shooting methods,  will fail under normal circumstances
because either (1) the integration tends to blow up to infinity when shooting 
from $x=0$ or (2) the integration is identically zero when shooting from $x=1$,
given the boundary conditions $y(1)=y'(1)=0$.  An embedded exponentially-fitted
Runga-Kutta method~\cite{rk} is reported to work.  But it is beyond the scope
of an undergraduate research project.  The relaxation method, on
the other hand, will in principle handle two singular boundary points.  In
this paper, we adapt the relaxation codes used in 
{\it Numerical Recipes}~\cite{nr}.  In order to establish a continuity
between this paper and the said reference, the same notations are used.

The basic idea of the relaxation method is to convert the differential 
equation into a finite difference equation.  Error functions are then
considered.  For instance, the error function at the $k$-th mesh point of the 
$j$-th differential equation (one of $N$ coupled differential equations)
\be
\left.{dy_{j}\over dx}\right|_{x_{k}}=g(x_{k},y_{1},\dots,y_{N})
\ee
is simply
\be
E_{j,k}=y_{j,k}-y_{j,k-1}-(x_{k}-x_{k-1})g(x_{k},x_{k-1},y_{1,k},y_{1,k-1},
\dots,y_{N,k},y_{N,y-1}).
\ee
The difference of $E_{j,k}$ is approximated by a first order Taylor series
expansion, such that
\be
\sum^{N}_{n=1}S_{i,n}\Delta y_{n,k-1}+\sum^{2N}_{n=N+1}S_{i,n}\Delta
y_{n-N,k}=-E_{i,k},
\label{sum}
\ee
where
\be
S_{i,n}={\partial E_{i,k}\over \partial y_{n,k-1}},\qquad
S_{i,n+N}={\partial E_{i,k}\over \partial y_{n,k}}.
\ee
The crux of this paper is to show how the matrix elements $S_{i,k}$ of the 
non-relativistic Schr\"{o}dinger equation are calculated and how to streamline
the relaxation codes.  The ranges of the sums in
Eq~[\ref{sum}] are split over $[1,N]$ and $[N+1,2N]$ because of the
peculiarity of the relaxation codes used in {\it Numerical Recipes}.
We will not delve into the details of the codes in this paper.  Readers are
encouraged to refer to {\it Numerical Recipes} for a full explanation.  As a
motivation, it suffices to say that the main idea of the relaxation method
is to begin with initial guesses of $y_{j,k}$ and relax them to the
approximately true values by calculating the errors $E_{i,k}$ to correct
$y_{j,k}$ iteratively.  $y_{j,k}$ are components of a solution vector in an 
$NM$-dimensional vector space.  Intuitively we can think of the relaxation 
process as rotating an initial vector into another vector under the 
constraints of $E_{i,k}$.  Since $y=0$ is a trivial solution, the relaxation
process has a tendency to diminish those components of the solution vector 
that correspond to the wavefunction and its derivative.  We will return to 
this point later in the paper.

Let $M-1$ be the number of mesh points and $h$ is step size, then
\be
h={1\over M-1},
\ee
and
\ba
x_{k}&=&(k-1)h,\\
y_{k}&\equiv&y(x_{k}).
\ea
We define the vector $y_i$ as
\begin{displaymath}
\left\{
\begin{array}{l l}
& y_{1} = y \\
& y_{2} = y' \\
& y_{3} = E
\end{array}
\right.
\end{displaymath}
such that
\ba
y_{1}' &=& y_{2} \label{y1p}\\
y_{2}' &=& -{2\over 1-x}\,y_{2}-{1\over (1-x)^{4}}\left[2\mu\,a_{0}^{2}
\left(y_{3}+{1-x\over x}\,{e^{2}\over a_{0}}\right)-\left({1-x\over x}
\right)^{2}l(l+1)\right]y_{1}
\label{y2p}\\
y_{3}' &=& 0 \label{y3p}
\ea
A first order finite derivative at $x_{k}$ is given as
\be
y_{k}'={y_{k}-y_{k-1}\over h}.
\ee
In the relaxation formalism, $E_{i,k}$ are to be zeroed.  In this case,
it is easy to see that Eq.~[\ref{y1p}] can be rewritten as
\be
E_{1,k}=y_{1,k}-y_{1,k-1}-{h\over 2}(y_{2,k}+y_{2,k-1}).
\ee
The corresponding $S_{1,j}$ $S$-matrix elements are
\ba
S_{1,1}&=&{dE_{1,k}\over dy_{1,k-1}}=-1,\\
S_{1,2}&=&{dE_{1,k}\over dy_{2,k-1}}=-{h\over 2},\\
S_{1,3}&=&{dE_{1,k}\over dy_{3,k-1}}=0,\\
S_{1,4}&=&{dE_{1,k}\over dy_{1,k}}=1,\\
S_{1,5}&=&{dE_{1,k}\over dy_{2,k}}=-{h\over 2},\\
S_{1,6}&=&{dE_{1,k}\over dy_{3,k}}=0.
\ea
Similarly, Eq.~[\ref{y2p}] is rewritten as
\ba
E_{2,k}&=&y_{2,k}-y_{2,k-1}+{2h\over 1-{x_{k}+x_{k-1}\over 2}}
({y_{2,k}+y_{2,k-1}\over 2}) 
+ {h\over \left(1-{x_{k}+x_{k-1}\over 2}\right)^{4}}\nonumber\\
&&\quad\times
\left[2\mu\,a_{0}^{2}\left({y_{3,k}+y_{3,k-1}\over 2}+{1-{x_{k}+x_{k-1}
\over 2}\over{x_{k}+x_{k-1}\over 2}}\,{e^{2}\over a_{0}}\right)
\right.\nonumber\\
&&\quad\qquad-\left.\left({1-{x_{k}+x_{k-1}\over 2}\over
{x_{k}+x_{k-1}\over 2}}\right)^{2}\,l(l+1)\right]
\,{y_{1,k}+y_{1,k-1}\over 2}.\label{e2k}
\ea
The associating $S_{1,j}$ $S$-matrix elements are
\ba
S_{2,1}&=&{dE_{2,k}\over dy_{1,k-1}}\nonumber\\
&=&{h\over 2\left(1-{x_{k}+x_{k-1}\over 2}\right)^{4}}
\left[2\mu\,a_{0}^{2}\left({y_{3,k}+y_{3,k-1}\over 2}+
{1-{x_{k}+x_{k-1}\over 2}\over{x_{k}+x_{k-1}\over 2}}\,{e^{2}\over a_{0}}
\right)\right.\nonumber\\
&&\qquad\qquad\qquad\qquad-\left.\left({1-{x_{k}+x_{k-1}\over 2}\over
{x_{k}+x_{k-1}\over 2}}\right)^{2}\,l(l+1)\right],\label{s21}\\
S_{2,2}&=&{dE_{2,k}\over dy_{2,k-1}}\nonumber\\
&=&-1+{h\over 1-{x_{k}+x_{k-1}\over 2}},\\
S_{2,3}&=&{dE_{2,k}\over dy_{3,k-1}}\nonumber\\
&=&{h\mu\,a_{0}^{2}\over \left(1-{x_{k}+x_{k-1}\over 2}\right)^{4}}\,
{y_{1,k}+y_{1,k-1}\over 2},\\
S_{2,4}&=&{dE_{2,k}\over dy_{1,k}}=S_{2,1},\\
S_{2,5}&=&{dE_{2,k}\over dy_{2,k}}\nonumber\\
&=&1+{h\over 1-{x_{k}+x_{k-1}\over 2}},\\
S_{2,6}&=&{dE_{2,k}\over dy_{3,k}}=S_{2,3}.
\ea
At last, Eq.~[\ref{y3p}] is rewritten as
\be
E_{3,k}=y_{3,k}-y_{3,k-1},
\ee
and the associating $S_{3,j}$ matrix elements are
\ba
S_{3,1}&=&{dE_{3,k}\over dy_{1,k-1}}=0,\\
S_{3,2}&=&{dE_{3,k}\over dy_{2,k-1}}=0,\\
S_{3,3}&=&{dE_{3,k}\over dy_{3,k-1}}=-1,\\
S_{3,4}&=&{dE_{3,k}\over dy_{1,k}}=0,\\
S_{3,5}&=&{dE_{3,k}\over dy_{2,k}}=0,\\
S_{3,6}&=&{dE_{3,k}\over dy_{3,k}}=1.
\ea
At the first boundary point $x=0$, we have $n_{1}=1$.  The boundary condition 
is $y(0)=0$.  It translates to
\be
E_{3,1}=y_{1,1}.
\ee
The only non-zero $S$-matrix element is
\be
S_{3,4}=1.
\ee
At the second boundary point $x=1$, we have $n_{2}=2$.  The boundary conditions
are $y(1)=y'(1)=0$, or equivalently
\ba
E_{1,M}&=&y_{1,M},\\
E_{2,M}&=&y_{2,M}.
\ea
The only non-zero matrix elements are
\ba
S_{1,4}&=&1,\\
S_{2,5}&=&1.
\ea

The hydrogen atom finite difference equation can be generalized to any Coulomb 
pentential by substituting $e^{2}\to\lambda_{C}$.  To adapt the code to solve 
NRSE with a linear potential, the only modifications needed are
\be
{1-x\over x}\,e^{2}\to-{x\over 1-x}\,\lambda_{L},
\ee
in Eq.~[\ref{y2p}],
\be
{1-{x_{k}+x_{k-1}\over 2}\over{x_{k}+x_{k-1}\over 2}}\,e^{2}\to
-{{x_{k}+x_{k-1}\over 2}\over 1-{x_{k}+x_{k-1}\over 2}}\,\lambda_{L}
\ee
in Eqs.~[\ref{e2k},\ref{s21}], and the omission of all $a_{0}$ from the
matrix elements.

\section{Exact Solution of the Hydrogen Atom}
The non-relativistic hydrogen atom is solved exactly by series expansion.
The first few radial wavefunctions are found to be~\cite{greiner}
\ba
R_{10}(r)&=&{1\over\sqrt{\pi}}\,\gamma_{1}^{3\over 2}r\,e^{-\gamma_{1}r},
\label{radial1}\\
R_{20}(r)&=&{1\over\sqrt{\pi}}\,\gamma_{2}^{3\over 2}r\,
(1-\gamma_{1}r)\,e^{-\gamma_{2}r},\\
R_{21}(r)&=&{1\over\sqrt{\pi}}\,\gamma_{2}^{5\over 2}r^{2}\,e^{-\gamma_{2}r},
\label{radial3}
\ea
where
\be
\gamma_{n}\equiv{1\over n\,a_{0}}
\ee
With the redefinition of variable by Eqs.~[\ref{redef1},\ref{redef2}] and
the omission of normalization constants, Eqs.~[\ref{radial1}--\ref{radial3}]
can be rewritten as
\ba
R_{10}(z)&\sim& z\,e^{-z},\label{rz1}\\
R_{20}(z)&\sim& z\,(1-z)\,e^{-{z\over 2}},\\
R_{21}(z)&\sim& z^{2}\,e^{-{z\over 2}}.\label{rz3}
\ea
Eqs.~[\ref{rz1}--\ref{rz3}] will be used to test the relaxation codes in
Section~[\ref{num_res}].

\section{Exact $S$-state Solution for the Linear Potential}
The $S$-state eigenvalue of the non-relativistic Schr\"{o}dinger equation with
a linear potential can be solved exactly in configuation space.
We shall use the analytic results to check our numerical results.  The
non-relativistic Schr\"{o}dinger equation can be written as
\begin{equation}
\left( {d^{2}\over dr^{2}} + {2\over r}\,{d\over dr} \right)\,R -
2\mu[\lambda_{L}\,r - E]\,R = 0. \label{linear_r}
\end{equation}
Let $S\equiv r\,R$, then Eq.~[\ref{linear_r}] can be simplified as
\begin{equation}
{d^{2}\over dr^{2}}S - 2\mu[\lambda_{L}\,r - E]S = 0.
\end{equation}
Define a new variable
\begin{equation}
x\equiv \left( {2\mu\over \lambda_{L}^{2}} \right)^{1\over 3}
[\lambda_{L}r - E], \label{linear_s}
\end{equation}
such that Eq.~[\ref{linear_r}] can be transformed as
\begin{equation}
S'' - xS = 0,
\end{equation}
which is the Airy equation.  The solution which satisfies the boundary
condition $S\to 0$ as $x\to \infty$ is the Airy function
${\rm Ai}(x)$.  It is easy to show that the eigen-energy formula is
\begin{equation}
E_{n} = -x_{n}\,\left( {\lambda_{L}^{2}\over 2\mu} \right)^{1\over 3},
\end{equation}
where $x_{n}$ is the $n$-th zero of the Airy function counting from
$x=0$ along $-x$.  
In reference~\cite{norbury91}, the values
$\lambda_{L} = 5$ and $\mu = 0.75$ are used.  In this case, the eigen-energy 
formula is
\begin{equation}
E_{n} = -2.554364772\,x_{n}.
\end{equation}

\section{Numerical Results}
\label{num_res}
The matrix elements in the routine \texttt{difeq} have been calculated in 
section~\ref{der}.  All but the main driver subroutines are left intact.  The 
intial guesses of the wavefunction and its derivative in the main driver
program are just the square of the sine function and its derivative.
The codes of the modified driver programs,\texttt{bohr.c} and 
\texttt{linear.c}, and the associating \texttt{difeq.c} are listed in the 
appendix of this paper.  All the codes are written in $C$.

For large number of mesh points ($1000\le N\le 10000$), the code generates 
smooth wavefunctions and energies which do not correspond to any eigen 
modes.  An explicit Schr\"{o}dinger solver is known to be 
conditionally unstable~\cite{iitaka,succi}.  This kind of instabilities 
motivates numerical schemes such as the Numrov method~\cite{numerov},
the symplectic scheme~\cite{symplectic}, microgenetic algoritm~\cite{micro},
and others~\cite{fast}.  For smaller number of mesh points ($40\le N\le 100$),
it is observed that the relaxed eigenvalues are largely similar to the initial 
guesses and that the smoothness of the relaxed eigenfunctions is very
sensitive to the initial guesses of the eigenvalue values.
Therefore we adopt the criterion that the smoothness of the relaxed 
wavefunction picks out an initial guess as the correct eigenvalue when the
number of mesh points is small enough.  This 
criterion has certain {\em a priori} appeal because we expect a real physical
wavefunction to be smooth.  The reason for using the initial guess instead of
the relaxed eigenvalue is based on the observation that the former is more
accurate than the latter when the code is calibrated using exact results.
Table~[\ref{comp}] illustrates the motivation of the smoothness criterion.
The combination of a small number of mesh points and the smoothness criterion 
is the simplest way to make the relaxation code workable as an explicit
Schr\"{o}dinger solver.

\begin{figure}[ht]
\begin{center}
\epsfig{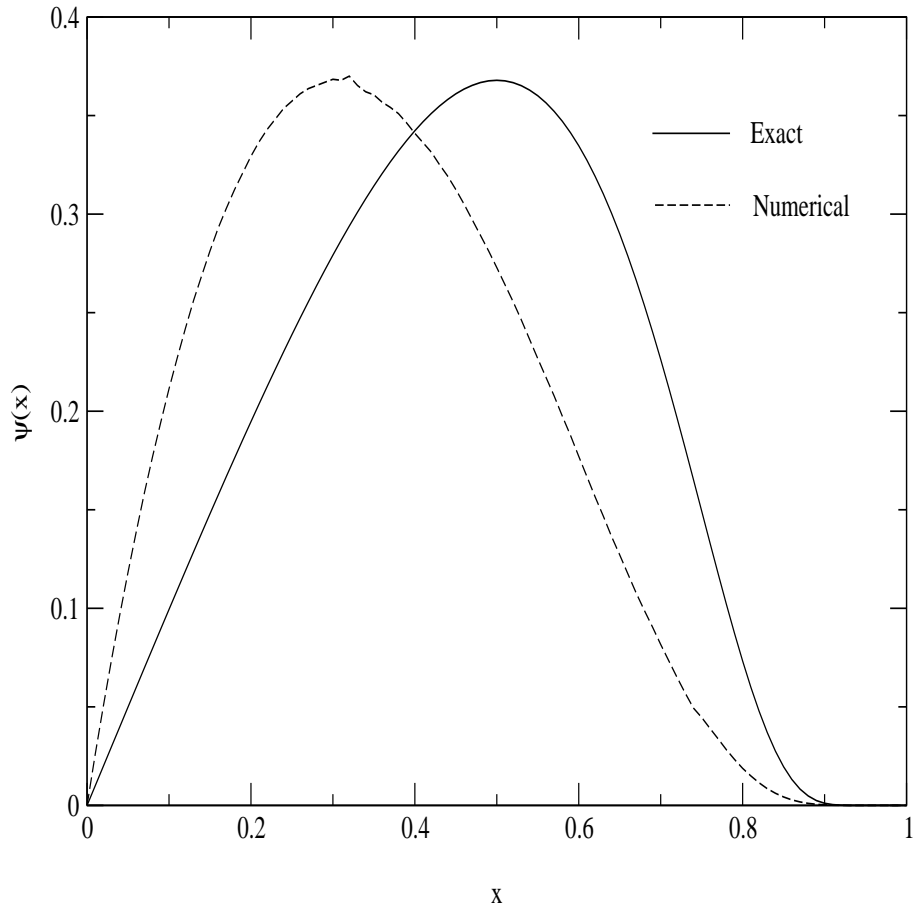}
\caption{\label{r10}
The relaxed $n=1$, $l=0$ wavefunction of the non-relativistic hydrogen atom
compared against the exact wavefunction.  The numerical eigenvalue is 
$-13.59827\,{\rm eV}$, versus the exact value of $-13.598289\,{\rm eV}$.}
\end{center}
\end{figure}

\begin{figure}[ht]
\begin{center}
\epsfig{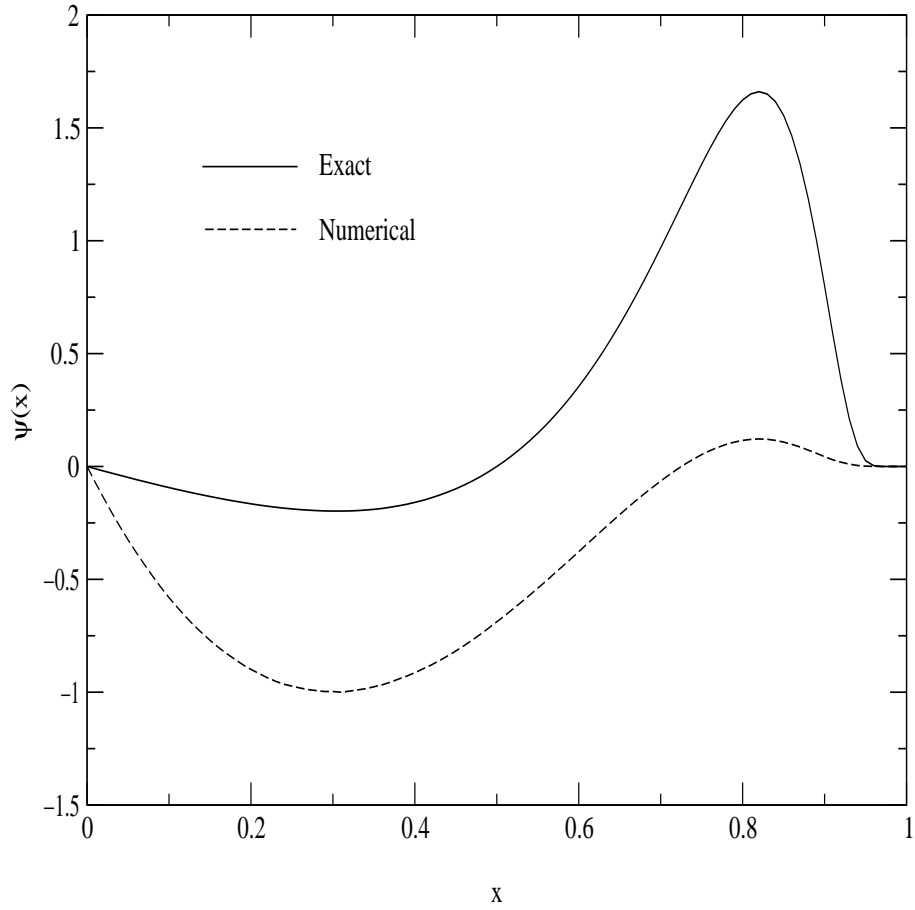}
\caption{\label{r20}
The relaxed $n=2$, $l=0$ wavefunction of the non-relativistic hydrogen atom
compared against the exact wavefunction.  The numerical eigenvalue is 
$-3.399750\,{\rm eV}$, versus the exact value of $-3.399572\,{\rm eV}$.}
\end{center}
\end{figure}

\begin{figure}[ht]
\begin{center}
\epsfig{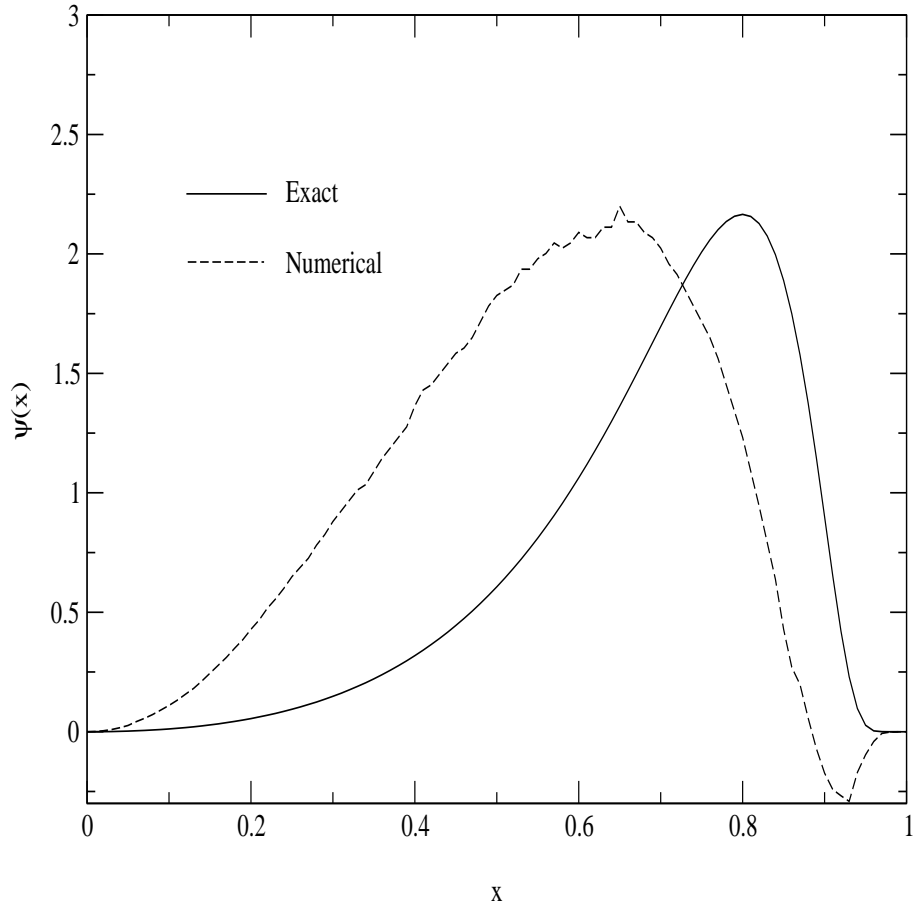}
\caption{\label{r21}
The relaxed $n=2$, $l=1$ wavefunction of the non-relativistic hydrogen atom
compared against the exact wavefunction.  The numerical eigenvalue is 
$-1.510056\,{\rm eV}$, versus the exact value of $-1.510921\,{\rm eV}$.}
\end{center}
\end{figure}

Figs.~[\ref{r10}--\ref{r21}] compare the relaxed and exact wavefunctions
of the hydrogen atom
for the first few quantum numbers.  The relaxed wavefunction typically has a 
vanishingly small amplitude.  It is explained by the tendency of the relaxation
routine to relax the wavefunction to the trivial solution $\psi=0$.  In order
to compare the relaxed and exact wavefunctions, the amplitude of the 
diminishing relaxed wavefunction is rescaled to match the exact wavefunction.  
Despite the large descrepancies between the relaxed and exact wavefunctions
of the hydrogen atom in Figs.~[\ref{r10},\ref{r20}], both the 
relaxed and exact wavefunctions share the same basic features.  Both relaxed 
wavefunctions satisfy the smoothness criterion.  However, there are 
some difficulties in obtaining a smooth relaxed wavefunction for 
Fig.~[\ref{r21}].  The relaxed wavefunction in this case also has an 
unwelcomed kink on the right hand side of the plot, which the exact 
wavefunction does not show.  Hence the relaxed and exact wavefunctions do not 
share the same features in this case.  The discrepancy here is mostly due to
degeneracy.  The eigen-energy formula of the hydrgon atom~\cite{greiner}
\be
E_{nl}=-{e^{2}\over 2\,a_{0}}\,{1\over (n+l)^{2}}
\ee
reveals the degeneracy over the $n$ and $l$ quantum states.  For example, 
$E_{21}=E_{30}$ such
that the wavefunctions of these two states are degenerate by virtue of
having the same eigenvalue.  In principle, any linear combination of the
degenerate solution vectors also satisfies the
finite difference equation Eq.~[\ref{de_a0}].  It also means that the relaxed
wavefunction can be any linear combination of the degenerate wavefunctions.
This problem will plague the numerical solutions of all degenerate states at 
higher $n$ and $l$.  Imperfect wavefucntions is not a liability when the
goal is to calculate the eigenvalues in the simplest possible way.
In this case, the relaxed wavefunction is used merely as a guide to pick out 
the correct eigenvalue and not as a final product.

\begin{figure}[ht]
\begin{center}
\epsfig{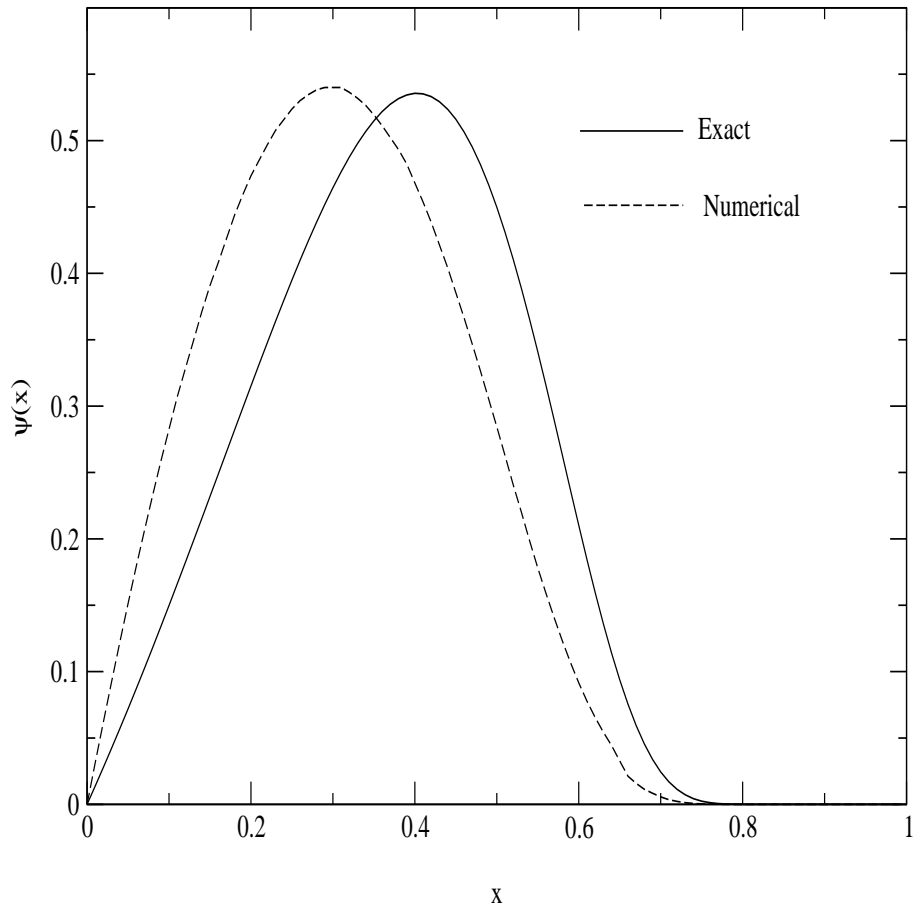}
\caption{\label{linear1}
The relaxed $l=0$, $n=1$ state wavefunction of a non-relativistic 
Schr\"{o}dinger equation with a linear potential using 
$\lambda_{L}=5\,{\rm GeV}$ and $\mu=0.75\,{\rm GeV}$.  The numerical 
eigenvalue is $5.9719\,{\rm GeV}$, versus the exact value of 
$5.972379\,{\rm GeV}$.  The relaxed wavefunction has been rescaled to
match the exact wavefunction.}
\end{center}
\end{figure}

\begin{figure}[ht]
\begin{center}
\epsfig{file=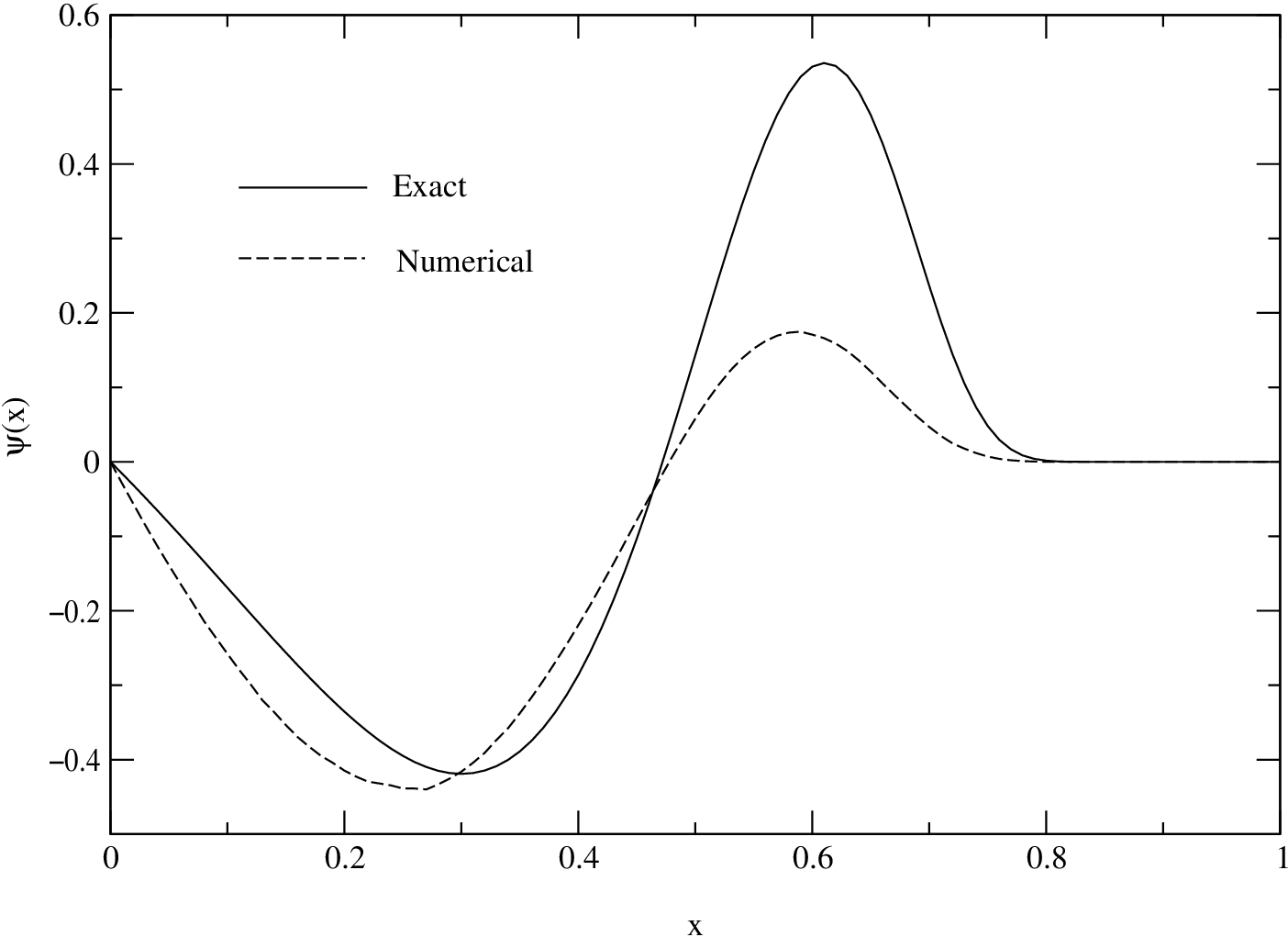,width=12cm,height=12cm}
\caption{\label{linear2}
The relaxed $l=0$, $n=2$ state wavefunction of a non-relativistic 
Schr\"{o}dinger equation with a linear potential using 
$\lambda_{L}=5\,{\rm GeV}$ and $\mu=0.75\,{\rm GeV}$.  The numerical 
eigenvalue is $10.441\,{\rm GeV}$, versus the exact value of 
$10.442114\,{\rm GeV}$.  The relaxed wavefunction has been rescaled to
match the exact wavefunction.}
\end{center}
\end{figure}

The situation of the numerical solution of the linear potential is slightly
better.  Figs~[\ref{linear1},\ref{linear2}] illustrate the relaxed 
wavefunctions of a linear potential compared to the exact wavefunctions.
The wavefunctions of a linear potential are not degenerate.  The
relaxation method is expected to be relatively successful in the absence
of degeneracy.  
In this paper, we use $\lambda_{L}=5\,{\rm GeV}$ and $\mu=0.75\,{\rm GeV}$.
Table~[\ref{lintab}] lists the eigenvalues of
the first few $l$ values in the case of the linear potential.

\section{Conclusion}
In this paper, we show how to circumvent the problem of conditional
instability associating with an explicit Schr\"{o}dinger solver when the
relaxation method is used.  The relaxation code with the combination of small
number of mesh points and the smoothness criterion gives good approximate 
eigenvalues in both the linear and Coulomb potential cases.  The role of
the relaxed wavefunction is simply to guide the selection of the correct
eigenvalue and is not used as an answer.  More accurate eigenvalues can be 
obtained by solving the momentum space integral equation 
using the Nystrom method~\cite{nystrom}.  In this work, we have solved only
the non-relativistic Schr\"{o}dinger equation with the linear and Coulomb
potentials.  In the case of a relativistic equation, the Hamiltonian contains 
a kinetic term $\sqrt{p^{2}+m^{2}}-m$ which is difficult to solve numerically 
with $r$-space codes.  The 
$p$-space solution using the Nystrom method on the other hand has the 
advantages of stability, accuracy and robustness over the $r$-space solution 
using the relaxation or shooting methods.  For these reasons, we prefer the
$p$-space codes over the $r$-space codes for production purposes.  
Nevertheless, the $r$-space code is useful in checking the $p$-space results 
whenever the former is available.  The numerical $r$-space calculation is a 
natural starting
point for students to learn scientific computation because they would have 
already seen the exact $r$-space solutions in a traditional quantum mechanics 
curriculum.  Since the numerical solution of the hydrogen atom is not readily 
available in publications, it is hoped that the numerical method presented
in this paper provides the simplest numerical scheme to supplement the 
discussion of the Schr\"{o}dinger equation in
an advanced undergraduate or beginning graduate quantum mechanics course.

\section{Acknowledgment}
We thank our advisor Prof. John W. Norbury for motivating this project.

\newpage
\section{Appendix}
\subsection{Non-relativistic Hydrogen Atom}
\texttt{bohr.c} is a driver program for solving the non-relativistic hydrogen
atom using the relaxation method.  The inputs are $n$ (the principal
quantum number), $l$ (the orbital angular
momentum quantum number), $E$ (the initial guess of the ground state 
eigen-energy of the hydrogen atom which is approximately -13.6) and
$scale$ (a scale factor to adjust the amplitude of the solution).  The output 
is a file \texttt{bohr.dat} which contains the relaxed wavefunction.

\begin{verbatim}
/* bohr.c */

#include <stdio.h>
#include <math.h>
#include "difeq.c"
#include "/recipes/c/solvde.c"
#include "/recipes/c/bksub.c"
#include "/recipes/c/pinvs.c"
#include "/recipes/c/red.c"
#define NRANSI
#include "/recipes/c/nrutil.h"
#include "/recipes/c/nr.h"
#include "/recipes/c/nrutil.c"
#define NE 3
#define M 101
#define NB 1
#define NSI NE
#define NYJ NE
#define NYK M
#define NCI NE
#define NCJ (NE-NB+1)
#define NCK (M+1)
#define NSJ (2*NE+1)
#define pi 3.1415926535897932384626433

int l,mpt=M;
float h,x[M+1],mu,e2,a0;

int main(void)  /* Program bohr.c */
{
	void solvde(int itmax, float conv, float slowc, float scalv[],
		int indexv[], int ne, int nb, int m, float **y, float ***c, 
		float **s);
	int i,itmax,k,indexv[NE+1],n;
	float conv,deriv,sinpx,cospx,q1,slowc,scalv[NE+1],eigen,guess;
	float **y,**s,***c,max,scale;
	FILE *fp;

	mu=0.5107208e6;
	e2=7.297353e-3;
	a0=1.0/mu/e2;
	y=matrix(1,NYJ,1,NYK);
	s=matrix(1,NSI,1,NSJ);
	c=f3tensor(1,NCI,1,NCJ,1,NCK);
	itmax=100;
	conv=1.0e-5;
	slowc=1.0;
	h=1.0/(M-1);
	printf("\nn, l, E, scale = ");
	scanf("%d%d%f%f",&n,&l,&guess,&scale);
	indexv[1]=1;
	indexv[2]=2;
	indexv[3]=3;
	for (k=1;k<=(M-1);k++) {
		x[k]=(k-1)*h;
		sinpx=sin((n-l)*pi*x[k]);
		cospx=cos((n-l)*pi*x[k]);
		y[1][k]=sinpx*sinpx;
		y[2][k]=2.0*(n-l)*pi*cospx*sinpx;
		y[3][k]=guess/(l+n)/(l+n);
	}
	x[M]=1.0;
	y[1][M]=0.0;
	y[3][M]=guess/(l+n)/(l+n);
	y[2][M]=0.0;
	scalv[1]=1.0;
	scalv[2]=1.0;
	scalv[3]=-guess/(l+n)/(l+n);
	solvde(itmax,conv,slowc,scalv,indexv,NE,NB,M,y,c,s);
	eigen=-13.598289/(n+l)/(n+l);
	printf("\nl\tE_in\t\tnumerical E\texact E\t\tnumerical-exact\n");
	printf("%d\t%f\t%f\t%f\t%f\n\n",l,guess/(l+n)/(l+n),
	       y[3][1],eigen,y[3][1]-eigen);
	fp=fopen("bohr.dat","w");
	max=0.0;
	for(i=1;i<=M;i++) if(fabs(y[1][i])>max) max=fabs(y[1][i]);
	for(i=1;i<=M;i++) fprintf(fp,"%f\t%e\n",(i-1)*h,y[1][i]/max*scale);
	fclose(fp);
	free_f3tensor(c,1,NCI,1,NCJ,1,NCK);
	free_matrix(s,1,NSI,1,NSJ);
	free_matrix(y,1,NYJ,1,NYK);
        return 0;
}
#undef NRANSI
\end{verbatim}

The \texttt{difeq.c} code below defines the $S$-matrix elements of the finite
difference equation of the non-relativistic hydrogen atom and is called by 
the relaxation subroutine \texttt{solvde.c}.

\begin{verbatim}
/* difeq.c for the hydrogen atom driver program */

extern int mpt,l;
extern float h,x[],mu,e2,a0;

void difeq(int k, int k1, int k2, int jsf, int is1, int isf, int indexv[],
	int ne, float **s, float **y)
{
  float xk,y1k,y2k,y3k;
	if (k == k1) {
	  s[3][3+indexv[1]]=1.0;
	  s[3][3+indexv[2]]=0.0;
	  s[3][3+indexv[3]]=0.0;
	  s[3][jsf]=y[1][1];
	} else if (k > k2) {
		s[1][3+indexv[1]]=1.0;
		s[1][3+indexv[2]]=0.0;
		s[1][3+indexv[3]]=0.0;
		s[1][jsf]=y[1][mpt];
		s[2][3+indexv[1]]=0.0;
		s[2][3+indexv[2]]=1.0;
		s[2][3+indexv[3]]=0.0;
		s[2][jsf]=y[2][mpt];
	} else {
		s[1][indexv[1]] = -1.0;
		s[1][indexv[2]] = -0.5*h;
		s[1][indexv[3]]=0.0;
		s[1][3+indexv[1]]=1.0;
		s[1][3+indexv[2]] = -0.5*h;
		s[1][3+indexv[3]]=0.0;
		xk=(x[k]+x[k-1])/2.0;
		y1k=(y[1][k]+y[1][k-1])/2.0;
		y2k=(y[2][k]+y[2][k-1])/2.0;
		y3k=(y[3][k]+y[3][k-1])/2.0;
		s[2][indexv[1]]=h/2.0/pow(1.0-xk,4.0)*
		  (2.0*mu*a0*a0*(y3k+(1.0/xk-1.0)*e2/a0)
		   -pow(1.0/xk-1.0,2.0)*l*(l+1));
		s[2][indexv[2]] = -1.0+h/(1.0-xk);
		s[2][indexv[3]]=h*mu/pow(1.0-xk,4.0)*y1k*a0*a0;
		s[2][3+indexv[1]]=s[2][indexv[1]];
		s[2][3+indexv[2]]=2.0+s[2][indexv[2]];
		s[2][3+indexv[3]]=s[2][indexv[3]];
		s[3][indexv[1]]=0.0;
		s[3][indexv[2]]=0.0;
		s[3][indexv[3]] = -1.0;
		s[3][3+indexv[1]]=0.0;
		s[3][3+indexv[2]]=0.0;
		s[3][3+indexv[3]]=1.0;
		s[1][jsf]=y[1][k]-y[1][k-1]-h*y2k;
		s[2][jsf]=y[2][k]-y[2][k-1]+2.0*h/(1.0-xk)*y2k+
		  h/pow(1.0-xk,4.0)*(2.0*mu*a0*a0*(y3k+(1.0/xk-1.0)*e2/a0)
				     -pow(1.0/xk-1.0,2.0)*l*(l+1))*y1k;
		s[3][jsf]=y[3][k]-y[3][k-1];
	}
}
\end{verbatim}

\newpage
\subsection{NRSE with a Linear Potential}
\texttt{linear.c} is a driver program for solving the non-relativistic
Schr\"{o}dinger equation with a linear potential using the relaxation
method.  The inputs are $n$ (the principal quantum number), $l$ (the orbital 
angular momentum quantum number), $E$ (the initial guess of
the eigen-energy) and $scale$ (a scale factor to adjust the amplitude of
the wavefunction)  The output is a file \texttt{linear.dat} which contains
the relaxed wavefunction.

\begin{verbatim}
/* linear.c */

#include <stdio.h>
#include <math.h>
#include "difeq.c"
#include "/recipes/c/solvde.c"
#include "/recipes/c/bksub.c"
#include "/recipes/c/pinvs.c"
#include "/recipes/c/red.c"
#define NRANSI
#include "/recipes/c/nrutil.h"
#include "/recipes/c/nr.h"
#include "/recipes/c/nrutil.c"
#define NE 3
#define M 101
#define NB 1
#define NSI NE
#define NYJ NE
#define NYK M
#define NCI NE
#define NCJ (NE-NB+1)
#define NCK (M+1)
#define NSJ (2*NE+1)
#define pi 3.1415926535897932384626433

int l,mpt=M;
float h,x[M+1],mu,lambdal;

int main(void)
{
	void solvde(int itmax, float conv, float slowc, float scalv[],
		int indexv[], int ne, int nb, int m, float **y, float ***c, 
		float **s);
	int i,itmax,k,indexv[NE+1],n;
	float conv,deriv,sinpx,cospx,q1,slowc,scalv[NE+1],max,e,scale;
	float **y,**s,***c;
	FILE *fp;

	mu=0.75;
	lambdal=5.0;
	y=matrix(1,NYJ,1,NYK);
	s=matrix(1,NSI,1,NSJ);
	c=f3tensor(1,NCI,1,NCJ,1,NCK);
	itmax=100;
	conv=1.0e-6;
	slowc=1.0;
	h=1.0/(M-1);
	printf("\nn, l, E scale = ");
	scanf("%d%d%f%f",&n,&l,&e,&scale);
	indexv[1]=1;
	indexv[2]=2;
	indexv[3]=3;
	for (k=1;k<=(M-1);k++) {
		x[k]=(k-1)*h;
		sinpx=sin(n*pi*x[k]);
		cospx=cos(n*pi*x[k]);
		y[1][k]=sinpx*sinpx;
		y[2][k]=2.0*n*pi*cospx*sinpx;
		y[3][k]=e;
	}
	x[M]=1.0;
	y[1][M]=0.0;
	y[3][M]=e;
	y[2][M]=0.0;
	scalv[1]=1.0;
	scalv[2]=1.0;
	scalv[3]=e;
	solvde(itmax,conv,slowc,scalv,indexv,NE,NB,M,y,c,s);
	fp=fopen("linear.dat","w");
	max=0.0;
	for(i=1;i<=M;i++) if (fabs(y[1][i])>max) max=fabs(y[1][i]);
	for(i=1;i<=M;i++) fprintf(fp,"%f\t%f\n",(i-1)*h,y[1][i]/max*scale);
	fclose(fp);
	printf("\nl\tE_i\t\tE_f\n");
	printf("%d\t%f\t%f\n\n",l,e,y[3][1]);
	free_f3tensor(c,1,NCI,1,NCJ,1,NCK);
	free_matrix(s,1,NSI,1,NSJ);
	free_matrix(y,1,NYJ,1,NYK);
        return 0;
}
#undef NRANSI
\end{verbatim}

The \texttt{difeq.c} code below define the $S$-matrix elements of the finite
difference equation of a non-relativistic Schr\"{o}dinger equation with
a linear potential and is called by the relaxation subroutine 
\texttt{solvde.c}.
 
\begin{verbatim}
/* difeq.c for the linear potential driver program */

extern int mpt,l;
extern float h,x[],mu,lambdal;

void difeq(int k, int k1, int k2, int jsf, int is1, int isf, int indexv[],
	int ne, float **s, float **y)
{
  float xk,y1k,y2k,y3k;
	if (k == k1) {
	  s[3][3+indexv[1]]=1.0;
	  s[3][3+indexv[2]]=0.0;
	  s[3][3+indexv[3]]=0.0;
	  s[3][jsf]=y[1][1];
	} else if (k > k2) {
		s[1][3+indexv[1]]=1.0;
		s[1][3+indexv[2]]=0.0;
		s[1][3+indexv[3]]=0.0;
		s[1][jsf]=y[1][mpt];
		s[2][3+indexv[1]]=0.0;
		s[2][3+indexv[2]]=1.0;
		s[2][3+indexv[3]]=0.0;
		s[2][jsf]=y[2][mpt];
	} else {
		s[1][indexv[1]] = -1.0;
		s[1][indexv[2]] = -0.5*h;
		s[1][indexv[3]]=0.0;
		s[1][3+indexv[1]]=1.0;
		s[1][3+indexv[2]] = -0.5*h;
		s[1][3+indexv[3]]=0.0;
		xk=(x[k]+x[k-1])/2.0;
		y1k=(y[1][k]+y[1][k-1])/2.0;
		y2k=(y[2][k]+y[2][k-1])/2.0;
		y3k=(y[3][k]+y[3][k-1])/2.0;
		s[2][indexv[1]]=h/2.0/pow(1.0-xk,4.0)*(2.0*mu*
					(y3k+xk/(xk-1.0)*lambdal)
					-pow(1.0/xk-1.0,2.0)*l*(l+1));
		s[2][indexv[2]] = -1.0+h/(1.0-xk);
		s[2][indexv[3]]=h*mu/pow(1.0-xk,4.0)*y1k;
		s[2][3+indexv[1]]=s[2][indexv[1]];
		s[2][3+indexv[2]]=2.0+s[2][indexv[2]];
		s[2][3+indexv[3]]=s[2][indexv[3]];
		s[3][indexv[1]]=0.0;
		s[3][indexv[2]]=0.0;
		s[3][indexv[3]] = -1.0;
		s[3][3+indexv[1]]=0.0;
		s[3][3+indexv[2]]=0.0;
		s[3][3+indexv[3]]=1.0;
		s[1][jsf]=y[1][k]-y[1][k-1]-h*y2k;
		s[2][jsf]=y[2][k]-y[2][k-1]+2.0*h/(1.0-xk)*y2k+
		  h/pow(1.0-xk,4.0)*(2.0*mu*(y3k+xk/(xk-1.0)*lambdal)
				     -pow(1.0/xk-1.0,2.0)*l*(l+1))*y1k;
		s[3][jsf]=y[3][k]-y[3][k-1];
	}
}
\end{verbatim}

\newpage

\newpage

\newpage

\newpage

\newpage

\newpage

\newpage

\newpage

\newpage

\begin{table}[ht]
\caption{Comparisons among the inital guesses, the relaxed and exact
eigenvalues of the non-relativistic hydrogen atom and the Schr\"{o}dinger 
equation with a linear potential using $\lambda_{L}=5\,{\rm GeV}$, 
$\mu=0.75\,{\rm GeV}$.}
\begin{center}
\vskip 10pt
\begin{tabular}{ccrrr}
\hline
&&&Coulomb&\\
$n$ & $l$ & Initial & Numerical & Exact \\
\hline
1 & 0 & -13.598270 & -13.621142 & -13.598289 \\
2 & 0 &  -3.399750 &  -3.400535 &  -3.399572 \\
2 & 1 &  -1.510056 &  -1.510060 &  -1.510921 \\
\hline\hline
&&&Linear&\\
$n$ & $l$ & Initial & Numerical & Exact \\
\hline
1 & 0 &  5.9719 &  6.146734 &  5.972379 \\
2 & 0 & 10.4410 & 10.418742 & 10.442114 \\
\hline
\end{tabular}
\end{center}
\label{comp}
\end{table}

\newpage

\begin{table}[ht]
\caption{The numerical eigenvalues of the non-relativistic 
Schr\"{o}dinger equation with 
a linear potential using $\lambda_{L}=5\,{\rm GeV}$, $\mu=0.75\,{\rm GeV}$
and $n=1$.}
\begin{center}
\vskip 10pt
\begin{tabular}{cr}
\hline
$l$ & Numerical \\
\hline
0 & 5.9719 \\
1 & 8.5850 \\
2 & 10.8514 \\
3 & 12.9020 \\
4 & 14.9790 \\
5 & 16.5845 \\
\hline
\end{tabular}
\end{center}
\label{lintab}
\end{table}

\end{document}